\documentclass[]{interact}

\usepackage{lineno}
\usepackage{hyperref}
\usepackage{xcolor}
\usepackage{float}
\usepackage{epstopdf}
\usepackage[caption=false]{subfig}

\usepackage{dsfont}

\usepackage{booktabs}

\usepackage[natbibapa,nodoi]{apacite}
\usepackage{xcolor}

\theoremstyle{plain}

\theoremstyle{definition}

\theoremstyle{remark}

\begin{document}

\articletype{Computers, Environment, and Urban Systems}

\title{Human vs. AI Safety Perception? Decoding Human Safety Perception with Eye-Tracking Systems, Street View Images, and Explainable AI}

\author{
\name{
Yuhao Kang\textsuperscript{a,b},
Junda Chen\textsuperscript{c},
Liu Liu\textsuperscript{b},
Kshitij Sharma\textsuperscript{d},
Martina Mazzarello\textsuperscript{b},
Simone Mora\textsuperscript{b,d},
Fábio Duarte\textsuperscript{b}\thanks{Corresponding author: Fábio Duarte. Email: fduarte@mit.edu},
and Carlo Ratti\textsuperscript{b}}
\affil{
\textsuperscript{a}GISense Lab, Department of Geography and the Environment, The University of Texas at Austin, Austin, TX, United States;\\
\textsuperscript{b}Senseable City Lab, Department of Urban Studies and Planning, Massachusetts Institute of Technology, Cambridge, MA, United States; \\
\textsuperscript{c}Department of Computer Science and Engineering, University of California San Diego, San Diego, CA, United States; \\
\textsuperscript{d}Department of Computer Science, Norwegian University of Science and Technology, Trondheim, Norway
}
}

\maketitle

\begin{abstract}
The way residents perceive safety plays an important role in how they use public spaces, and it informs city planning and public policy. In recent years, studies have combined large-scale street view images and advanced computer vision techniques to measure the perception of safety of urban environments. Despite their success, such studies have often overlooked the specific environmental visual factors that draw human attention and trigger people's feelings of safety perceptions.
In this study, we introduce a computational framework that enriches the existing body of literature on place perception by using eye-tracking systems with street view images and deep learning approaches. 
Eye-tracking systems quantify not only what users are looking at but also how long they engage with specific environmental elements. 
This allows us to explore the nuance of which visual environmental factors influence human safety perceptions.
We conducted our research in Helsingborg, Sweden, where we recruited volunteers outfitted with eye-tracking systems. 
They were asked to indicate which of the two street view images appeared safer.
By examining participants' focus on specific features using Mean Object Ratio in Highlighted Regions (MoRH) and Mean Object Hue (MoH), we identified key visual elements that attract human attention when perceiving safe environments.
For instance, certain urban infrastructure (e.g., stairways and signboards) and public space (flags and chairs) features draw more human attention while the sky is less relevant in influencing safety perceptions.
These insights offer a more human-centered understanding of which urban features influence human safety perceptions.
Furthermore, we compared the real human attention from eye-tracking systems with attention maps obtained from eXplainable Artificial Intelligence (XAI) results. 
Several XAI models were tested, and we observed that XGradCAM and EigenCAM most closely align with human safety perceptual patterns.
Our framework provides a valuable approach to enhance the interpretability and trustworthiness of XAI models by comparing them with empirically observed human behavior data.
This study demonstrates the limitations of previous place perception studies that solely rely on street view images and computer vision techniques, which may not comprehensively capture the nuanced human experiences and behaviors at place.
The inclusion of technologies such as eye-tracking not only deepens our comprehension of human subjective experiences, but also contributes to the development of safer environments and communities.
\end{abstract}


\section{Introduction}
Understanding human safety perceptions in the built environment offers key perspectives for creating safe places, and fostering a sense of security and well-being among residents \citep{un2012enhancing, raco2007securing, ceccato2013moving}.
In alignment with the United Nations' Sustainable Development Goals (SDG), particularly the aim of fostering safe and inclusive cities and settlements \citep{abubakar2019prospects}, it is crucial to explore which places people perceive as safe and unsafe, and what urban features influence their perception beyond addressing criminal activities \citep{tabrizian2018exploring,rodriguez2020inclusive, brymer2021exploring, gobster2004human,rahm2021evening, kang2023understanding, ceccato2015aim}.
Historical perspectives in this field have yielded a variety of theories that illuminate the nature of the safe-environment nexus. 
The \textit{Broken Windows} theory, for instance, suggests that visible signs of disorder may lead to increased fear of crime \citep{o2019looking, doran2005investigating, wilson1982broken}.
Similarly, the concept of \textit{Defensible Space and Crime Prevention Through Environmental Design} emphasizes the significance of architectural and environmental design in deterring crime and promoting a sense of security \citep{jeffery1971crime, newman1973defensible}.  
Additionally, the \textit{Prospect-Refuge} theory suggests that environments offering clear visibility and places to hide or retreat are perceived as safer, influencing human behaviors and experiences choices \citep{appleton1975experience,ramanujam2006prospect}.
Consequently, an investigation of the subjective human sense of safety could provide novel insights into urban place-making and ultimately inform policies, foster safer communities and enhancing residents' sense of belonging.

To measure human place perceptions, recent studies have combined street view images and Artificial Intelligence (AI), particularly deep learning-based computer vision approach to measure human safety perceptions \citep{kang2023understanding, ito2024understanding, hamim2024towards, wang2019using, ramirez2021measuring, kang2023assessing, zhang2021perception, zhou2025using}.
This approach utilizes large volumes of street view images coupled with advanced computer vision techniques to examine the complex relationships between human safety perceptions and their environmental contexts.
The underlying hypothesis is that individuals' subjective responses to these street view images can accurately represent their perceptions of real-world environments.
Notably, this approach offers a more efficient and cost-effective alternative to traditional methods such as surveys and interviews \citep{biljecki2021street, kang2020review}.
However, despite its success, this approach faces two potential limitations in capturing the relationships between human safety perceptions and the built environment \citep{ito2024understanding}.

First, when analyzing the associations between human safety perceptions and objects within street view images, a notable issue refers to the undifferentiated treatment of all detected objects.
Researchers tend to model the safety-environment nexus merely by considering the percentage of the presence of objects in these images \citep{zhang2018measuring, ramirez2021measuring, larkin2021predicting, ogawa2024evaluating, dong2023assessing}.
However, such approaches may not accurately reflect true human behaviors, and may instead resemble mathematical manipulation rather than performing behavioral analysis.
This method may overlook the fact that, when observing these images, individuals often concentrate their attention selectively on certain objects rather than perceiving all objects as equally important \citep{uttley2018eye, he2024human}. 
Some smaller but unique objects in urban environments may draw more attention than larger, more commonplace ones.
Furthermore, even when two images have similar proportions of environmental objects, they may evoke significantly different impressions due to their diverse environmental settings, which influences human perceptions.
A larger proportion of urban elements in images does not necessarily indicate their significance in shaping human perceptual or behavioral responses. 
Human experiments remain essential to capture and reflect how individuals interact with their surrounding environments.
Thus, such a discrepancy highlights the critical need to enrich prior studies with a more nuanced, human-centered perspective on environmental perceptions \citep{kang2025human}.


Second, a research gap arises from the ``black box'' nature of deep learning in measuring human place perception studies.
Recently, the ethical implications of using artificial intelligence (AI) have attracted increasing attention, highlighting the need to critically assess the trustworthiness and potential biases in AI-generated outcomes \citep{jobin2019global,wach2023dark, kang2024artificial}.
In prior studies that utilized street view images and computer vision techniques, there have been efforts employing eXplainable Artificial Intelligence (XAI) methods such as Grad-CAM \citep{selvaraju2017grad} to delineate factors contributing to safety perceptions \citep{li2022measuring, moreno2021understanding, sangers2022explainability}.
The hypothesis is that these XAI methods can accurately reflect human behaviors and perceptions. 
Existing studies suggested that environmental features like greenery, cars, and sidewalks are important in influencing urban perceptions \citep{sangers2022explainability}. 
Despite these advancements, disparities highlighted by \citet{li2022measuring} between the outcomes of XAI-based methods and actual human perceptual preferences raise questions regarding the reliability of these metrics: 
To what degree do these measures accurately reflect human perceptions, and can we trust these metrics in understanding subjective human experiences?
Consequently, a deeper exploration of the associations between human subjective safety perceptions and tangible elements in urban environments is essential.
It is crucial to have a better understanding of the XAI methods utilized in prior studies by integrating more human perspectives.

To bridge the aforementioned two research gaps, we propose a computational framework that incorporates eye-tracking systems to understand how visual environmental factors trigger human safety perceptions.
Eye-tracking systems, an emerging technology, have been used in a variety of spatial cognition and environmental psychology studies, offering unique opportunities to capture human visual attention and perceptions \citep{kiefer2017eye, hollander2018seeing, dong2014eye}. 
The emergence of eye-tracking systems allows us to precisely track where and how long individuals focus their gaze when viewing images or environments \citep{goldberg1999computer, he2023geospatial}.
Inspired by prior studies that leverage psychological methods to understand human perceptions \citep{qin2023perceptions, yang2024urban}, we integrate eye-tracking systems to enrich our understanding of place perceptions.
In our research, the use of eye-tracking systems could help identify specific features that attract an individual's attention.
Such an approach is crucial to align more closely with real-world human behaviors and reflect human mental space which could more accurately identify and assess elements that influence safety perceptions \citep{shaw2021understanding, yang2025neurocognitive, hei2025leveraging}.
Furthermore, eye-tracking data, which more accurately reflect human true behaviors, might be utilized to validate the explainability of deep learning models, providing a potential alternative to unlocking the ``black box'' of these models.
We aim to validate the XAI-generated (eXplainable Artificial Intelligence) results by aligning them with eye-tracking results to help evaluate the robustness and trustworthiness of the workflow that combines street view images and computer vision techniques in measuring safety perceptions.
Consequently, integrating eye-tracking systems into place perception studies presents a promising avenue that will not only fill the existing research gaps but also gain a comprehensive understanding of the characteristics and reliability of methods in place perception studies.

To this end, this study aims to deepen our understanding of human safety perceptions by using emerging datasets and technologies including eye-tracking systems, street view images, and XAI. 
We aim to investigate the influence of environmental features in street view images on human safety perceptions.
Specifically, we ask the following research questions:\\
(1) What environmental features do individuals mostly focus on when assessing safety perception in cities? \\
(2) To what extent can eye-tracking system-based gaze heatmaps validate the outputs of XAI methods to enhance the understanding of their interpretability and trustworthiness for modeling safety perceptions?\\
To answer these questions, we first surveyed participants in Helsingborg, Sweden, to evaluate their perception of safety based on street view images of this city.
Participants were equipped with eye-tracking systems to evaluate their safety perceptions of street view images, allowing us to further record and analyze their gaze patterns.
After that, we generated a series of heatmaps to reflect their gaze patterns and identified environmental objects that drew significant attention.
Two metrics were developed, the Mean Object Ratio in Highlighted Regions (MoRH) and the Mean Object Hue (MoH) to detect important urban features.
We further compared the heatmaps from the eye-tracking systems and those generated by XAI methods such as Grad-CAM to understand their similarities and discrepancies.
Finally, we provided crucial insights for future place-making and urban design that contribute to the creation of safer environments, and discussed potential model biases to inform future technological advancements in the field.

The major contributions and innovations of this study are twofolds:
First, by integrating multiple emerging technologies such as eye-tracking systems, street view images, and XAI, this study provides a deep understanding of how environmental factors shape safety perceptions at places;  
We identified specific objects in urban landscapes that may trigger human safety perceptions and delineated how individuals visually engage with their surroundings.
This analysis provides practical guidance for environmental and urban design to enhance the sense of safety in communities.
Second, our study validates the results obtained from XAI with real-world eye-tracking data on safety perceptions and identified the XAI model that most closely aligns with human perceptions;
By doing so, our study offers implications for enhancing the trustworthiness of XAI models and advocates for enriching human-centered perspectives in the development of ethical AI methods to inform future urban planning and geography.

\section{Conceptual Framework and Preliminary}
\subsection{Conceptual Framework}
\begin{figure}[h]
    \centering
    \includegraphics[width=\textwidth]{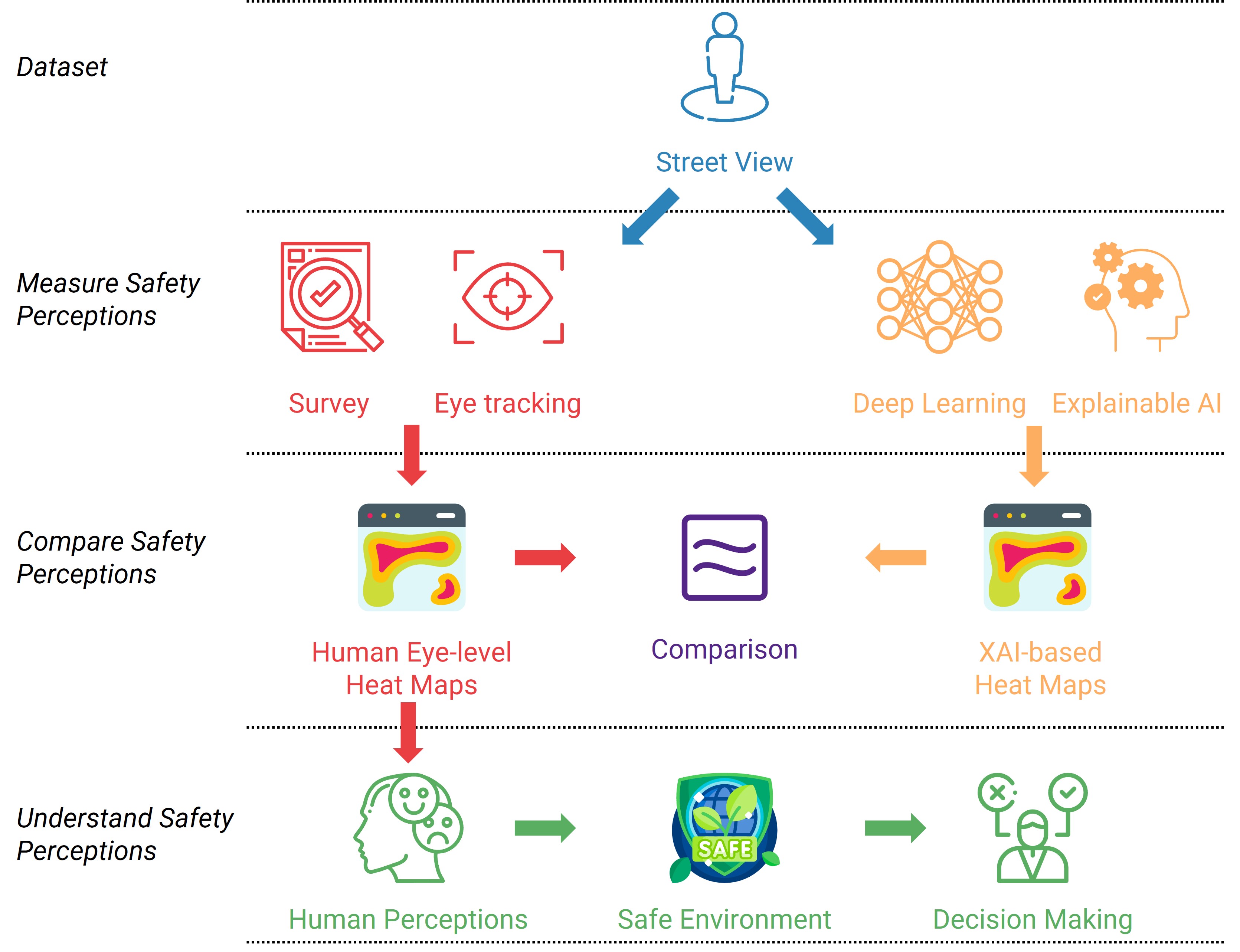}
    \caption{Conceptual framework of this study. We start from collecting street view images to measure perceived safety. Participants complete a survey and equip eye-tracking systems to produce human attention heatmaps. We also run deep learning-based models to predict safety scores with Explainable AI (XAI) approaches to generate XAI-based heatmaps. By comparing the two results, we aim to deepen our understanding of urban safe environments to inform decision-making for urban planning and design.}
    \label{fig:framework}
\end{figure}

Figure \ref{fig:framework} shows the overall computational framework of this study.
First, to collect human safety perceptions of the built environment, we established a survey based on a sample dataset of street view images.
We recruited local volunteers in Helsingborg to collect their safety perceptions.
They were outfitted with eye-tracking systems when they engaged with the urban landscapes in street view images.
Their behavioral data collected was then used to generate human attention heatmaps, which indicate regions within the images that attract significant human focus.
Second, following the collection of safety perception data, we applied image segmentation techniques to street view images to detect urban features.
These features were then correlated with the human attention heatmaps generated by eye-tracking systems \citep{yang2024urban}.
To facilitate a more nuanced analysis, we introduced two metrics including Mean Object Ratio in Highlighted Regions (MoRH) and Mean Object Hue (MoH), which allowed us to identify urban features that significantly capture human attention, thereby enabling us to assess their impact on safety perceptions effectively. 
After that, we employed a well-established workflow that integrates street view images with computer vision techniques to measure safety perceptions.
In particular, we leveraged eXplainable Artificial Intelligence (XAI) models and generated XAI-based heatmaps that highlight the importance of urban features while measuring safety perceptions using AI. 
By comparing these human attention heatmaps and XAI-based heatmaps, we offer valuable human insights regarding the trustworthiness of safety perceptions measured in prior studies to better align with human behaviors. 
It is worth noting that this study focuses on safety perceptions, and the proposed framework can be applied to other perceptual dimensions widely studied in urban planning such as liveliness and beauty.

\subsection{Preliminary: Measuring Safety Perceptions using Street View Images and Computer Vision}
Before introducing the methodologies employed in this study, here, we outline a common practice in prior place perception studies that combines street view images and computer vision, which forms the foundation of our approach.
In this study, we will leverage two distinct models that have been utilized to measure place perceptions, each based on a different dataset. 
The first model was developed based on the MIT Place Pulse dataset, which provides a global perspective on place perceptions with over 80,000 participants \citep{dubey2016deep}, hereafter referred to as the \textit{global dataset} and the \textit{global model}. 
The second model was constructed from a dataset with over 2,000 residents in Stockholm, Sweden \citep{kang2023assessing}, which aligns with the country in this paper, subsequently referred to as the \textit{Sweden dataset} and the \textit{Sweden model}.
The Sweden model delivers insights from a contextual perspective in Sweden, as prior studies have indicated that datasets derived from local people, with contextual knowledge, offer a more accurate reflection of place perceptions than the global dataset \citep{kang2023assessing, yao2019human}.
Combining both models could help provide a preliminary estimation of the safety perceptions from street view images from global and local perspectives.



The development of both models began with the launch of a survey inviting participants to evaluate their environmental perceptions through a set of selected street view images.
In these surveys, participants were presented with two randomly selected images and asked to select one of them to respond to survey questions such as ``Which place looks safer?''
Notably, the MIT Place Pulse dataset collected six dimensions of human place perceptions including safe, beautiful, depressing, lively, wealthy, and boring, while the Sweden dataset collected safety perceptions of the environment.
After collecting participants' preferences, we derived and calculated safety perceptual scores from these evaluations as indicators of the participants' impressions of each street view image.
These scores were then leveraged as proxies to represent people’s general safety perceptions of the surrounding environment.
Following this, we trained a Deep Convolutional Neural Network (DCNN) model to decipher the human perception of a place.
Upon completion of its training, the model could be utilized to estimate the perceptual score of any new street view image.
The human safety perceptual scores were delineated on a scale ranging from 1 to 9, where a higher score indicates a safer environment.
Both models were further leveraged to prepare the safety perception dataset, which will be utilized to measure human safety perceptions using eye-tracking systems. 
To further compare the attention maps generated from eye-tracking systems and XAI models, we performed those XAI methods based on the Sweden model, as it may better reflect residents' perceptions in Sweden \citep{kang2023assessing}.

\section{Measure Safety Perceptions with Eye-tracking Systems}
\subsection{Dataset}
\label{section:dataset}
We recruited participants and conducted our experiments in Helsingborg, Sweden. 
To ensure privacy, no identifiable personal information was collected from participants. 
A street view imagery dataset that includes 300 images was then constructed to represent the urban streetscapes of Helsingborg.
Street view imagery captures detailed characteristics of urban environments, from architectural styles and building facades to road networks and green spaces \citep{kang2020review,biljecki2021street}.
Given that street view imagery offers a comprehensive visual representation of urban landscapes, it has been used for collecting human safety perceptions under the hypothesis that human reactions to these street view images can serve as effective proxies for their perceptions of the built environment \citep{zhang2021perception,kang2023assessing}.
To create this dataset, we randomly selected images across the city to ensure the representativeness and spatial coverage. 
We also selected images that evoke people safe and unsafe perceptions to support consequent data collection experiments that integrates eye-tracking systems.
By leveraging eye-tracking systems to collect human reactions to the street view imagery, we aim to uncover a more nuanced understanding of how the urban environment influences residents' perception of safety.
More detailed data preprocessing steps for preparing the survey are provided in Appendix \ref{Appendix:dataset}

\subsection{Survey Setup}
\label{section:survey}
Based on the street view imagery dataset which contains 300 sample images, we launched a local online survey in Helsingborg and integrated eye-tracking systems to collect and understand safety perceptions \citep{kang2023assessing}.
We deployed eye-tracking systems in four places in Helsingborg. 
We utilized two models of eye-tracking systems, including the TobiiX3-120, offering 0.5 degrees of precision at 120 Hz, and the Tobii-4c, also with 0.5 degrees of precision but at 60 Hz, across a standard distance of 70 cm. Both devices were operated under a researcher's license, ensuring high-quality data collection. The data from the TobiiX3-120 was downsampled to 60Hz to standardize the output. 
Using Tobii's proprietary software, blinks were automatically filtered out, and both fixations and saccades were accurately identified through the integrated iVT-Tobii filter.

We adopted eye-tracking systems as a core place perception assessment tool in this study to complement traditional survey methods.
Eye-tracking systems could capture participants' visual attention when looking at images, offering insights into how individuals visually engage with different urban elements.
Unlike traditional methods such as questionnaires or semantic differential scales, which rely on subjective recall and post-hoc self-reporting, eye-tracking systems capture human gaze behaviors, enabling the measurement of more nuanced safety perceptions.
We invited passersby to share their safety perceptions in response to street view images.
Overall, responses from 127 participants were collected from this survey. 
On average, each street view image was compared 5.8 times with others.
We analyzed the demographic information of our participants. 
In total, we received valid responses from 127 participants.
The gender ratio (male vs. female) is 1.25:1, with White individuals comprising over 93\% of the population.

\begin{figure}[h]
    \centering
    \includegraphics[width=\textwidth]{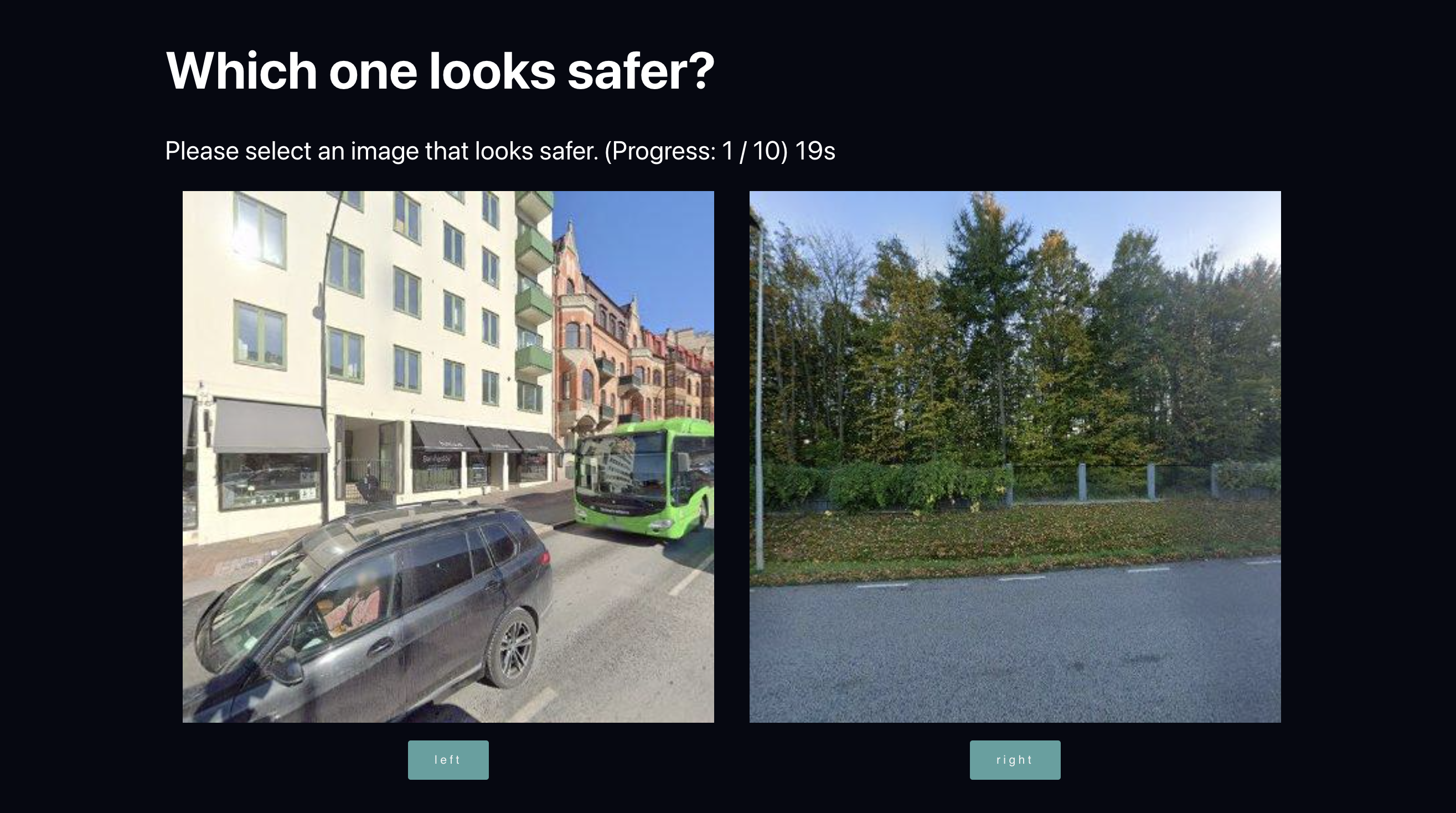}
    \caption{A user interface screenshot of the survey.}
    \label{fig:survey}
\end{figure}

Figure \ref{fig:survey} shows the user interface of the created survey. 
Each participant was first asked to provide their demographic details, including age and gender, for further analysis of potential population biases. 
Then, participants were exposed to ten pairs of two random street view images and asked to select, for each pair, ``Which place looks safer?''.
Each participant was equipped with an eye-tracking system so that their gazes when viewing the street view images were recorded.
By tracking participants' eye movements, we could identify which objects within the images attracted their attention, triggered their decisions, and therefore shaped their safety perceptions.
We followed the established workflows of surveys from prior studies to ensure consistency in our comparisons between human eye-level safety perceptions using eye-tracking systems, and AI-generated safety perceptions with XAI methods \citep{dubey2016deep, zhang2018measuring, kang2023assessing}.


\subsection{Human Attention Heatmaps}
The eye-tracking data, collected from participants in evaluating safe environments, was then translated as visual heatmaps.
These visual heatmaps obtained from eye-tracking systems are further termed \textit{human attention} heatmaps.
Visual heatmaps have been utilized as a common method for visualizing human attention through eye-tracking systems, as they effectively illustrate the focal areas observed by participants \citep{raschke2014visual, kiefer2017eye}. 
Eye-tracking systems capture the duration and locations of gaze fixations across street view images.
Based on this, we aggregated all gaze points of each pixel and performed a cumulative distribution function to represent the accumulated focus intensity for that pixel to create the heatmap \citep{raschke2014visual}.
It should be noted that this study only leveraged aggregated spatial heatmaps of images to capture patterns of visual attention across images.
Temporal aspects of gaze behavior were not modeled in this analysis.
As illustrated in Figure \ref{fig:heatmap}, the heatmaps use a color gradient to indicate the density of gaze attention: warmer colors such as reds and yellows highlight areas of interest (AOIs) that attracted more observation, whereas cooler colors represent AOIs that received less visual attention. 
Specifically, our analysis utilizes the hue scale as a metric for this spectrum of attention.
While raw hue values range from 0 to 360 degrees, we applied a linear rescaling transformation to map this spectrum onto a 0–150 range purely for visualization purposes. 
Regions with lower hue values (red/yellow, warmer colors) reflect high visual attention, and regions with higher hue values (e.g., blue, cooler colors) reflect lower human attention.
This transformation did not affect any computational analyses and data distributions, and facilitates the intuitive interpretation of gaze intensity and attention distribution across street view images.
It also enables us to further identify features in urban streetscapes that potentially impact human safety perceptions.
Moreover, it allows for a comparative analysis between heatmaps generated from eye-tracking systems and those generated using XAI models.

After translating images into heatmaps to represent human visual attention, we further categorize street view images into two groups based on participant selection frequencies in the pairwise safety comparison task.
Specifically, we classified images into a ``more frequently selected as safer" group (safe) and a ``less frequently selected as safer" group (unsafe).
Images classified as \textit{safe} were those consistently chosen as safer at least three times, whereas \textit{unsafe} images were those not chosen as safer images at least three times.
These labels are derived from aggregated human choices, rather than from model scores, and reflect behaviorally evoked safety perceptions during comparison.
By doing so, we could further link these behavior-grounded safety perceptions with urban features that draw human attention, thereby shedding light on modeling human-environmental interactions.

\subsection{Image Segmentation for MoR}
We further extracted objects from street view images using image segmentation methods, aiming to analyze what elements and objects may trigger human safety perceptions.
To do so, we leveraged a cutting-edge deep learning method based on Vision Transformers (ViT), utilizing the Dense Prediction Transformer (DPT) model. 
The DPT model is trained on the ADE20K dataset and could identify 150 common objects in urban places \citep{ranftl2021vision, zhou2017scene}.
This image segmentation pipeline has been widely applied in recent place perception studies \citep{ma2025perceived, ceccato2025makes}.
As illustrated in Figure \ref{fig:heatmap}, urban features were identified with different colors in the example street view image.
We can quantify their proportions in street view images using the following equation:
\begin{equation}
    R_{o} = \frac{\sum_{i=1}^{h} \sum_{j=1}^{w} \mathds{1}(p{(i,j)} = o)}{h \times w}
\end{equation}
where $R_{o}$ indicates the proportion of the object $o$ within an image, and $p{(i,j)}$ refers to the pixel value at the row $i$ and column $j$ of the image.
It is computed by dividing the number of pixels corresponding to object $o$ by the total number of pixels in the street view images ($h$ and $w$ are the height and width of images).
By computing the mean values of $\overline{R_{o}}$ (Mean Object Ratio, MoR) across all images for each object, we could measure the frequency of urban features in urban landscapes captured in the street view imagery.

\subsection{Linking Attention Heatmaps for Important Feature Identification using MoRH and MoH}
Performing image segmentation to identify urban elements from street view images and build associations between urban features and perceptions has been a common practice in existing literature \citep{larkin2021predicting, zhang2018measuring, ramirez2021measuring}.
However, as illustrated in Figure \ref{fig:heatmap}, when viewing street view images, participants may have different focus and attention on certain objects.
Therefore, it is necessary to take people's focus into account.
To quantify what objects are important in shaping and affecting human safety perceptions, we propose two metrics: Mean Object Ratio in Highlighted Regions (MoRH, $\overline{RH_{(t, o)}}$) and Mean Object Hue (MoH, $\overline{Hue_{o}}$).

The first index, \textit{Mean Object Ratio in Highlighted Regions} (MoRH, $\overline{RH_{(t, o)}}$), is calculated to quantify the frequency of objects that appear in the most visually attention AOIs of an image, as defined by a specified threshold $t$.
Different from the Mean Object Ratio (MoR), which considers the average presence of objects throughout the entire image, the Mean Object Ratio in Highlighted Regions (MoRH) specifically focuses on those areas receiving the highest levels of human attention, thereby quantifying the appearance of objects in these AOIs.
Considering the variance in the size of these highlighted AOIs, the MoRH metric is normalized by the number of pixels of these AOIs within the image.
The MoRH is expressed as follows:
\begin{equation}
    RH_{(t, o)} = \frac{\sum_{i=1}^{h} \sum_{j=1}^{w} \mathds{1}(p{(i,j)} = o \wedge Hue(p{(i,j)}) \leq t))}{\sum_{i=1}^{h} \sum_{j=1}^{w} \mathds{1}(Hue(p{(i,j)})\leq t)}
\end{equation}
where $RH_{(t, o)}$ indicates the proportion of the object $o$ within the highlighted areas, characterized by hue values falling below the threshold $t$.

The second index, \textit{Mean Object Hue (MoH)}, calculates the average hue value of a certain object in street view images.
By using the hue value to depict the intensity of human visual focus, the MoH value could indicate the level of attention directed toward a specific object.
Compared with MoR, MoH takes hue values as weights to indicate the importance of different urban features.
In our study, the hue values in heatmaps range from 0 to 150.
It is worth noting that we adjusted the MoH by computing $150 - MoH$ to get its reverse meaning for easier interpretation.
This ensures that a higher MoH, like 150, corresponds to a strong level of attention, and 0 represents a low level of attention to a particular object. 
By proposing the two metrics mentioned above, including MoRH and MoH values, we could quantify the importance of objects in street view images when participants perceive the images, rather than treating them equally.

\section{Understand Safety Perceptions}
To discover what street elements trigger human safety perceptions, we first analyzed street features by comparing the proportions of objects in street view images with Mean Object Ratio (MoR) and solely in highlighted areas using Mean Object Ratio in Highlighted Regions (MoRH).
After that, we identified important street elements based on Mean Object Hue (MoH).

\subsection{Object Detection without Eye-tracking Systems}
\label{obj_detection_without_eye-tracking}
\begin{figure}[h]
    \centering
    \includegraphics[width=\textwidth]{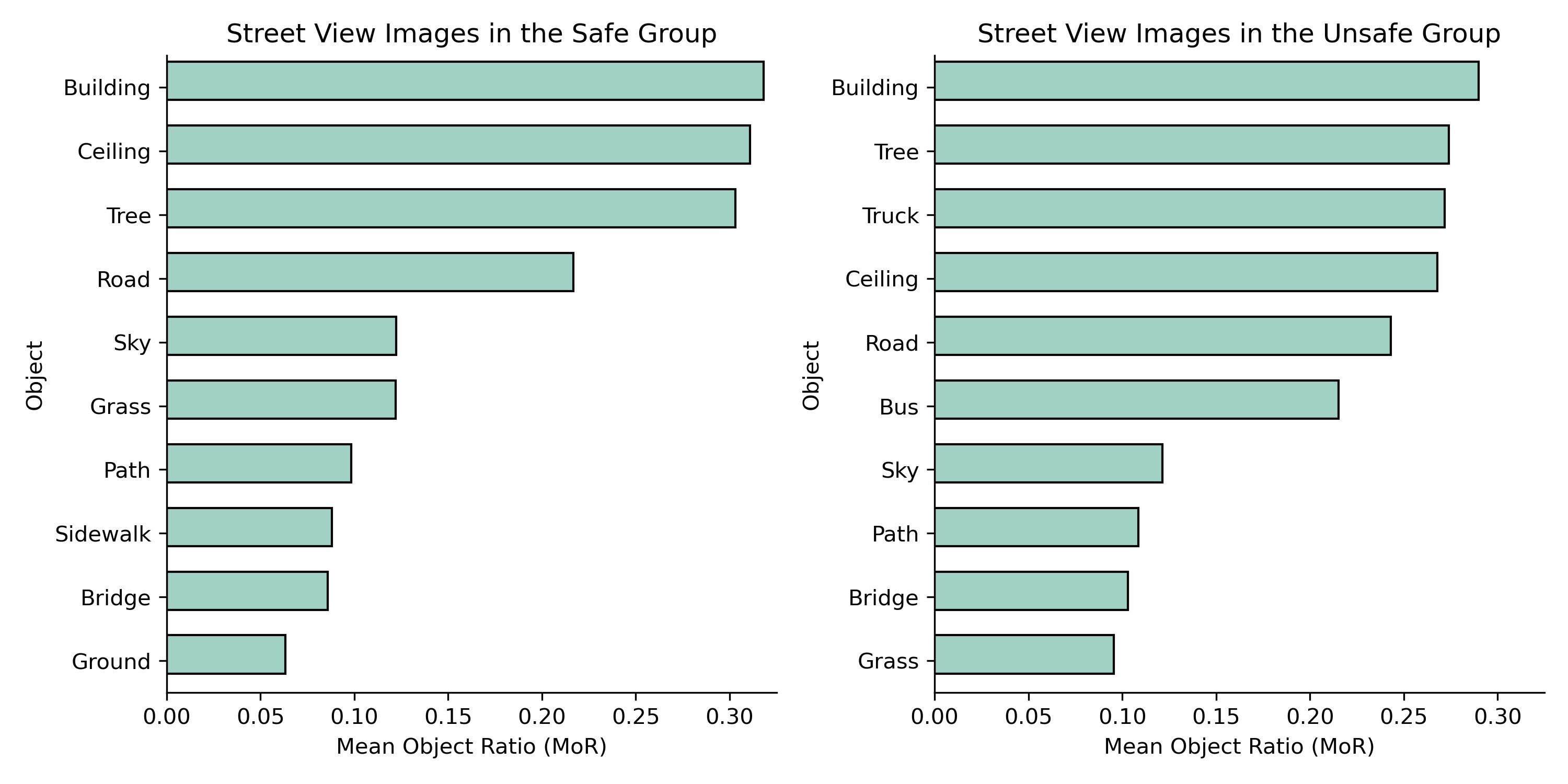}
    \caption{Top 10 objects based on Mean Object Ratio (MoR) in the two groups of street view images.}
    \label{fig:mor}
\end{figure}

Adopting methodologies utilized in prior studies, we employed image segmentation to quantify the occurrence of various objects in street view images.
We aim to reproduce prior practices to assess perceived urban safety without integrating eye-tracking systems. 
Given that we have divided street view images into two groups based on participant selection frequency: safe and unsafe, we identified the top 10 objects most commonly present in each group, as depicted in Figure \ref{fig:mor}.
Five key categories of street elements were identified, each contributing significantly to the composition of urban streetscapes:
\textit{Architectural features} such as buildings, ceilings, and bridges are frequently observed in street view images for creating urban streetscapes.
\textit{Pathways} including roads, paths, and sidewalks, are commonly seen in street view images, with sidewalks having more appearances in ``safe images'', indicating a potential positive correlation between pedestrian pathways and sense of safety.
\textit{Urban greenery}, such as trees and grass, appearing frequently in street view images, highlights the significance of green spaces in urban settings.
Variables such as sky and ground that influence the \textit{openness} and the overall atmosphere of urban scenes are also key elements in street view images.
In addition, the presence of \textit{vehicles} such as trucks and buses, often observed in ``unsafe images'', implies a possible negative association between vehicles and safe environments.
Overall, these observations are consistent with several findings from prior studies.
For example, elements such as buildings, trees, and sky typically occupy substantial proportions of pixel space in street view imagery \citep{ramirez2021measuring, li2022measuring}.
Urban greenery and open spaces are typically associated with positive environmental perceptions \citep{zhang2018measuring, ramirez2021measuring}.
While the presence of vehicles might have a negative impact on safe environments.
It is worth noting that the occurrence of street elements in images does not necessarily directly reflect their impacts on safety perceptions.
It emphasizes the complex human-environment relationships, highlighting the significance of delving into such interactions to better understand how various tangible street elements influence human subjective safety perceptions.

\begin{figure}[h]
    \centering
    \includegraphics[width=\textwidth]{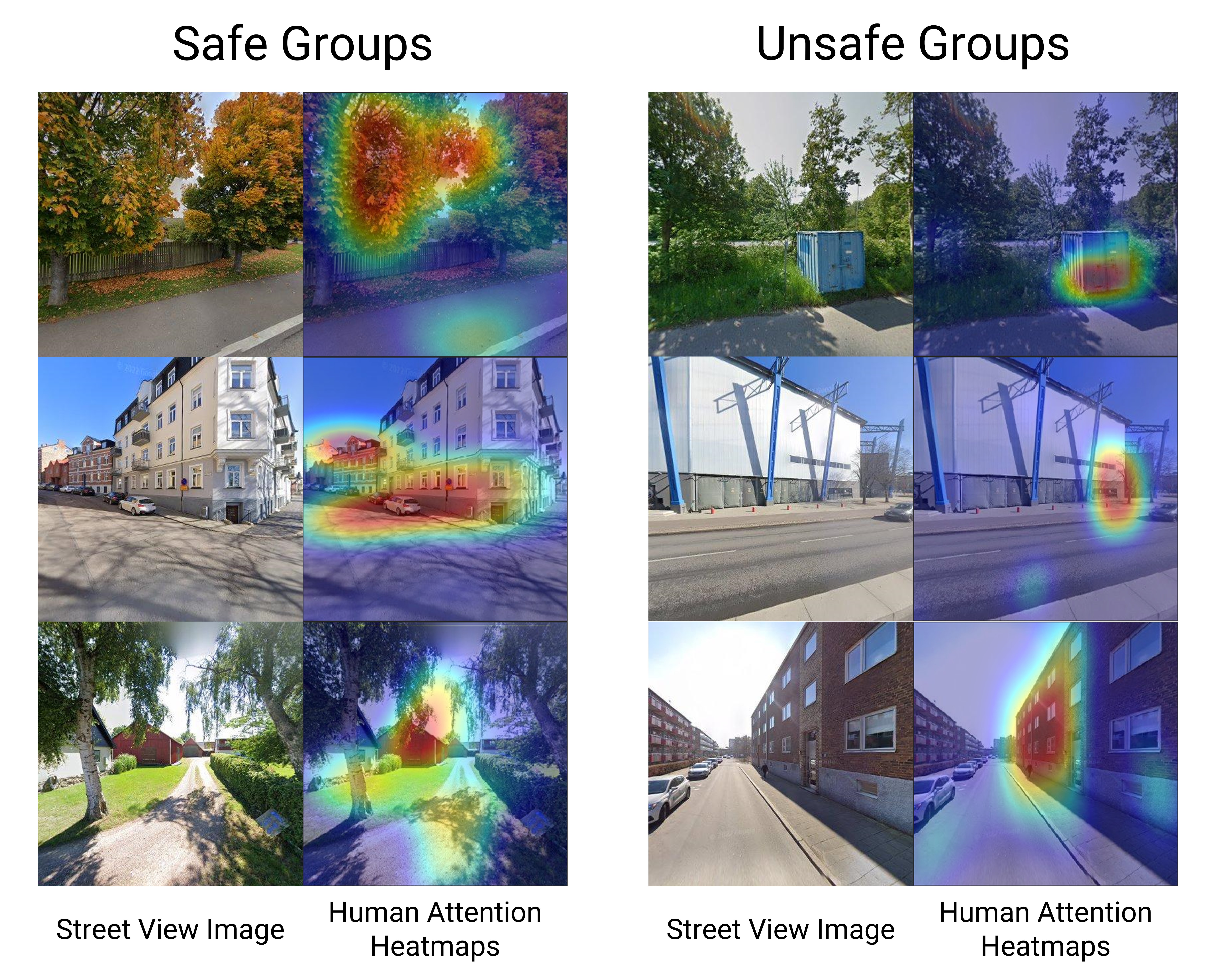}
    \caption{Sample street view images in safe and unsafe groups with aggregated human attention heatmaps.}
    \label{fig:sample}
\end{figure}

\subsection{Object Detection with Eye-tracking Systems}
By incorporating eye-tracking systems into our study, we can detect specific urban elements that attract human attention when observing street view images. 
Figure \ref{fig:sample} presents several representative images from the safe and unsafe groups together with aggregated human attention heatmaps from participants.
This overlay of attention maps supports a more nuanced understanding of the human behaviors and attention that shape urban safety perceptions.
To associate human gazes with urban elements, we performed image segmentation and computed the Mean Object Ratio in Highlighted Regions (MoRH) and Mean Object Hue (MoH) to identify important urban elements.

\begin{figure}[h]
    \centering
    \includegraphics[width=\textwidth]{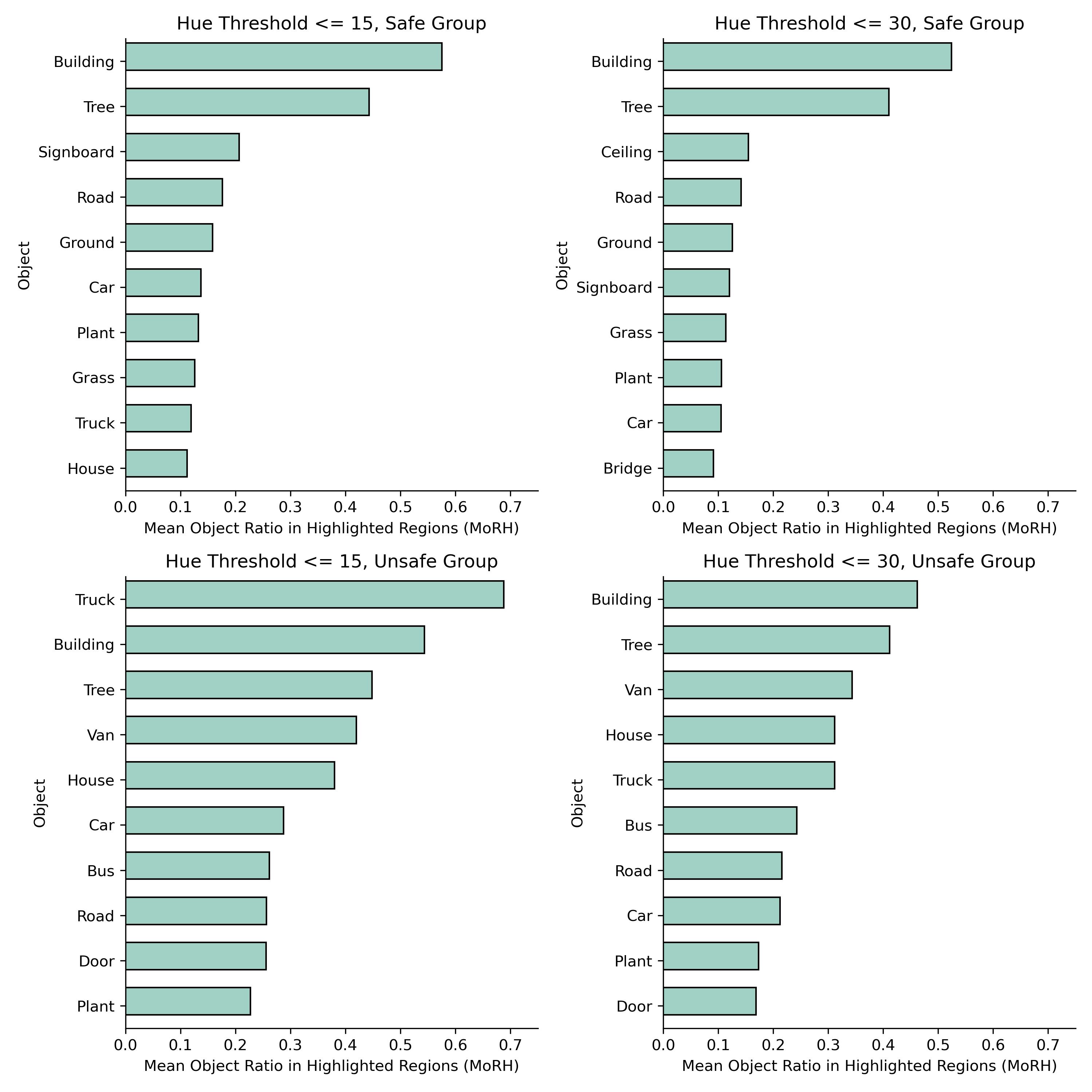}
    \caption{Top 10 objects based on Mean Object Ratio in Highlighted Regions (MoRH) in the two groups of street view images.}
    \label{fig:morh}
\end{figure}

In our analysis, it is critical to acknowledge that the previous results in Section \ref{obj_detection_without_eye-tracking} have given equal weights to all objects within street view images, potentially overshadowing the relative importance of some objects in attracting human attention. 
Figure \ref{fig:morh} illustrates the top 10 objects ranked by their MoRH in the two image groups (safe vs. unsafe).
The frequency of streetscape objects is calculated by performing image segmentation on the highlighted regions.
We evaluated two thresholds $t$, 15 and 30, to compute the MoRH, as we do not have a precise threshold. 
We then analyzed the most common street elements in those highlighted regions.

Comparing the results in Section \ref{obj_detection_without_eye-tracking}, we found notable discrepancies between the outcomes generated by eye-tracking systems and those obtained without this technology. 
For images in the safe group, signboards, cars, trucks, and plants were identified as more significant based on eye-tracking data which aligns with prior studies \citep{dong2023assessing}.
In terms of key street elements associated with images in the unsafe group, we found that trucks, vans, cars, houses, doors, and plants are more significant compared to results obtained without eye-tracking systems.
Conversely, ceilings were not among the top 10 objects in areas with the most human attention when $MoRH(t=15)$ in safe images, yet they were ranked among the top three objects when $MoRH(t=30)$.
This implies that they take up a considerable portion of the image but may not be the most important features in influencing perceived safe environments. 
Additionally, the eye-tracking data showed that the sky and pathways, such as paths and sidewalks, were perceived as less significant in both groups of images.
Such interesting findings demonstrate that sky may not be a focal variable in shaping human safety perceptions, different from prior work \citep{ogawa2024evaluating}.
This comparison provides a detailed analysis of how humans visually perceive and prioritize urban elements using eye-tracking technology.
It emphasizes the importance of exploring the nature of human visual perceptions rather than solely focusing on the occurrence of objects in images.

\begin{figure}[h]
    \centering
    \includegraphics[width=\textwidth]{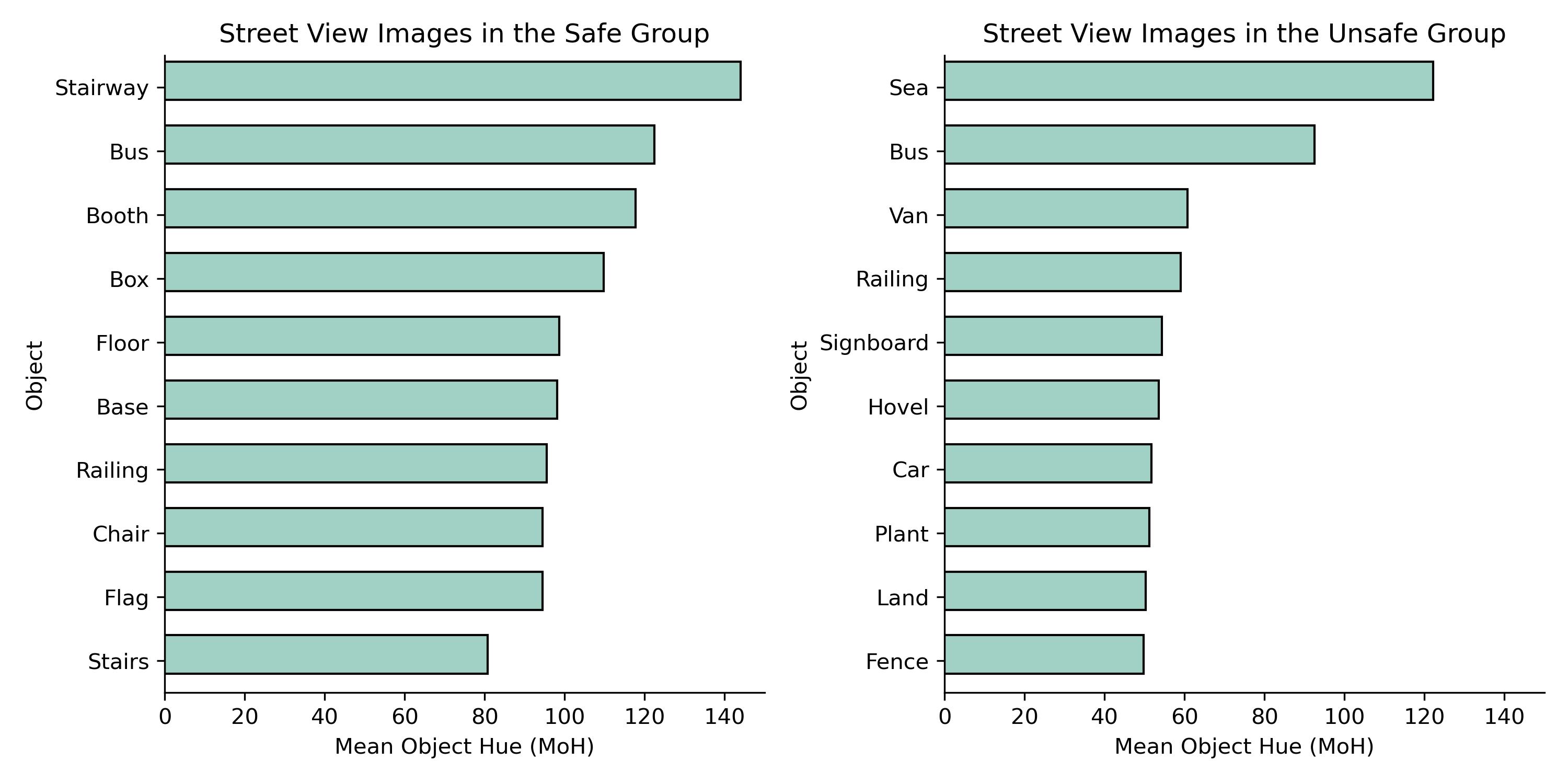}
    \caption{Top 10 objects based on Mean Object Hue (MoH) in the two groups of street view images.}
    \label{fig:moh}
\end{figure}

Next, we computed the Mean Object Hue (MoH) to analyze the attention intensity at the element level, considering the varying sizes of urban objects and elements.
Figure \ref{fig:moh} presents the top 10 objects ranked by MoH in images from both safe and unsafe groups. 
We observed significant discrepancies in results when human attention was considered in identifying important street elements, which highlights the complex relationship between physical urban elements and subjective human perceptions.
\textit{Transportation vehicles} such as buses, vans, and cars attracted more human focus, particularly in unsafe images, suggesting an association with unsafe environments.
\textit{Urban infrastructure} elements draw more human attention to both safe and unsafe images. 
Stairways, floors, bases, and railings, exhibit a positive correlation with safety perceptions, highlighting their role in creating secure urban places.
while elements like railings, signboards, fences, and hovels might be associated with negative safety perceptions, indicating potential barriers in urban environments.
\textit{Public space} features such as booths, chairs, flags, plants, and boxes also have high MoHs, and might be positively associated with safety perceptions, reflecting their roles in enhancing the quality and usability of public spaces.
Land and sea, as \textit{natural elements}, have shown negative associations with safety perceptions.
This aligns with prior work suggesting that open natural elements can evoke a sense of vulnerability due to their lack of enclosure, restricted visibility, and human presence \citep{nasar1997landscapes, stamps2005enclosure}. 
Environments with limited opportunities for surveillance, reduced escape routes, and isolated environments might be associated with increased perceptions of unsafety \citep{appleton1975experience,fisher1992fear}.
The results mentioned above not only demonstrated the crucial role of delving into human perceptions with eye-tracking systems but also offer new insights into illuminating the nature of the safe-environment nexus.

\section{Safety Perception Comparison}
This section aims to assess whether place perceptions measured using XAI methods could accurately reflect human perceptions.
Thus, we compare heatmaps generated by XAI techniques with those produced from human eye-tracking data.


\subsection{Explainable Artificial Intelligence (XAI)}
In prior studies, researchers have leveraged XAI methods such as Gradient-weighted Class Activation Mapping (GradCAM) to examine the associations between safety perceptions and built environments \citep{li2022measuring, moreno2021understanding}.
These methods help understand which elements in urban environments play important roles in shaping human subjective place perceptions.
In our study, we also followed this strategy and utilized Class Activation Mapping (CAM) to identify objects in street view images that significantly influenced the classification outcomes from the deep learning model.
In particular, we utilized CAM to create heatmaps from the Sweden model, as it may better reflect Swedish safety perceptions.

\begin{figure}[h]
    \centering
    \includegraphics[width=\textwidth]{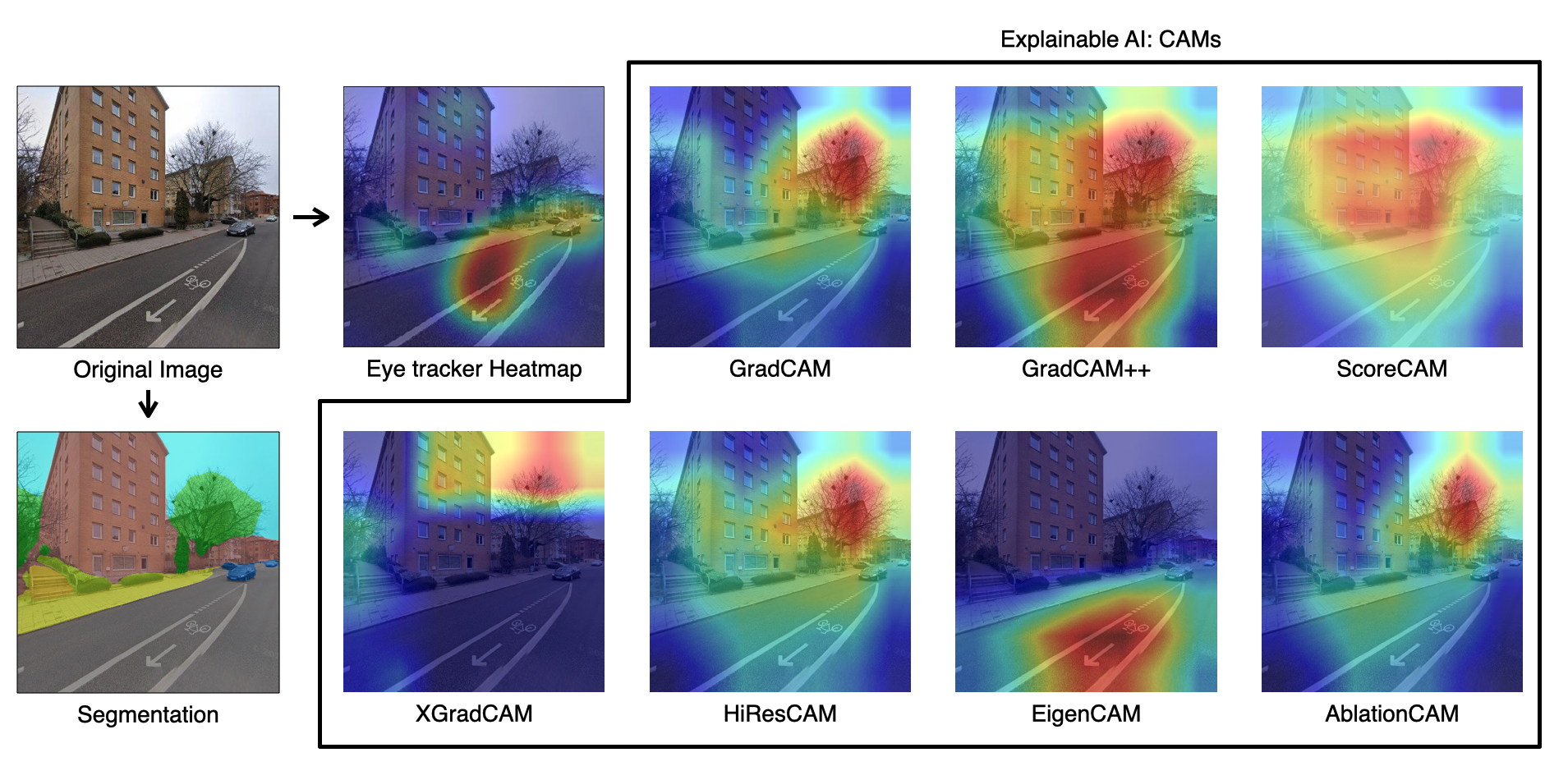}
    \caption{Examples of the original street view images, image segmentation results, human visual attention, and a series of heatmaps.}
    \label{fig:heatmap}
\end{figure}

Deep Convolutional Neural Networks (DCNNs) contain multiple convolutional layers. 
Each of these layers produces feature maps that capture different patterns of the input image, such as edges, textures, or shapes. 
These feature maps are essential as they highlight the specific characteristics that the model identifies, and are thereby utilized by CAM approaches to provide explanatory insights.
In particular, we leveraged seven CAM techniques, including GradCAM, ScoreCAM, GradCAMPlus, AblationCAM, XGradCAM, EigenCAM, and HiResCAM.
Each of these offers nuanced visualizations of influential regions within the input image, represented as heatmaps.
These approaches vary in their focus and characteristics.
GradCAM emphasizes the gradients flowing into the last convolutional layer;
ScoreCAM recalculates class scores, considering individual feature maps;
GradCAMPlusPlus extends GradCAM by considering the positive and negative impacts of the feature maps;
AblationCAM ablates each feature map and measures the output change;
XGradCAM introduces a modified formulation to produce more visually sharp regions;
EigenCAM leverages principal component analysis;
and HiResCAM focuses on providing high-resolution maps. 
The derived heatmaps showcase areas that have a substantial impact on the classification of safety perceptions.
The presence of warm hues (ranging from vibrant yellows to intense reds) within heatmaps indicates regions where the observed urban elements are more important in shaping subjective safety perceptions.
An example of these heatmap outputs can be seen in Figure \ref{fig:heatmap}, demonstrating the practical application and visual impact of our selected CAM techniques.
Since we have already obtained the human attention heatmaps, to enable comparison, we standardized the hue values of these XAI-based heatmaps between 0 and 150 as well.
After that, both groups of heatmaps were input into the image segmentation model to identify important objects and to further compare their differences.

\subsection{Image Similarity Comparison}
\label{image_similarity}
After we obtained the two groups of heatmaps, including human attention heatmaps and XAI-based heatmaps, we quantified their similarities to help understand the reliability and trustworthiness of XAI models and identified which one aligns most closely with human vision.
To accomplish this, we comprehensively measured the image similarity at two levels: \textit{scene} level and \textit{element} level \citep{kang2020review}.

At the scene level, we aim to quantify the overall similarity between any two images.
Two metrics, the L2 loss, and the Learned Perceptual Image Patch Similarity (LPIPS) score \citep{zhang2018unreasonable, jang2024place}, were computed between the two groups of heatmaps.
Both metrics range from 0 to 1 and assess the overall similarity between images. 
A lower L2 loss value indicates a greater similarity between human and XAI heatmaps; while lower LPIPS values indicate greater similarity between human and XAI heatmaps.
By using these two measures, we measured the image similarity between the human attention and XAI-based heatmap images to identify the XAI model that best matches human visual attention.
More technical details about these two metrics are provided in Appendix \ref{Appendix:similarity}.

At the element level, we focused on verifying whether the objects that received people's attention in both sets of heatmaps were consistent.
To accomplish this, we first calculated the MoH value for each object across all heatmaps. 
Each heatmap is depicted as a 150-dimensional vector.
We then compare the similarity between vectors of human attention heatmaps and their corresponding seven XAI-based heatmaps by calculating the cosine similarity.
The hypothesis is that objects that appeared in both heatmaps should be similar in two groups of images.
By doing so, we could determine which specific XAI model most aligns with human perceptions, offering valuable insights into the effectiveness and reliability of different XAI models when understanding human safety perceptions. 

\subsection{Comparison Results}



To answer the second research question, we evaluated the trustworthiness of using street view images and deep learning to measure safety perceptions at both scene and element levels.
At the scene level, we compared the image similarities between heatmaps generated by XAI models and those derived from human assessments utilizing eye-tracking systems.
Both the L2 loss function and LPIPS scores were computed.
The comparative analysis of these XAI models using the two metrics in Table \ref{tab:imgsimilarity} demonstrates similar trends. 
XGradCAM and EigenCAM have the lowest L2 loss values of 0.3285 and 0.3512, respectively.
They also ranked in the top 3 models with the lowest LPIPS scores of 0.5739 and 0.5478 (the lowest), meaning they have high similarities with heatmaps based on eye-tracking systems.
This indicates that the outputs from these two XAI approaches trained based on the Stockholm model closely match the results from using human eye-tracking systems. 
If we treat the human eye-tracking results as the benchmark for measuring human safety perceptions, these findings suggest that XGradCAM and EigenCAM may offer greater reliability and might be more trustworthy when using XAI models for future applications.
It also emphasizes the significance of involving humans in the process of developing and evaluating explainable artificial intelligence to derive interpretable insights.

\begin{table}[h]
\caption{Image similarities between human attention heatmaps and seven XAI-based heatmaps based on three measures. Top 2 most similar XAI methods are in bold.}
\label{tab:imgsimilarity}
\begin{tabular}{@{}llll@{}}
\toprule
Model Name     & Loss \(\downarrow\) & LPIPS Score \(\downarrow\) & Cosine Similarity \(\uparrow\) \\ \midrule
AblationCAM     & 0.4232 & 0.5855 & 0.0132 \\ 
EigenCAM        & \textbf{0.3512} & \textbf{0.5478} & \textbf{0.0143} \\
GradCAM         & 0.4357 & 0.5896 & 0.0130 \\
GradCAMPlusPlus & 0.4998 & 0.5891 & 0.0100 \\
HiResCAM        & 0.4386 & \textbf{0.5688} & 0.0127 \\
ScoreCAM        & 0.4631 & 0.5806 & 0.0100 \\
XGradCAM        & \textbf{0.3285} & 0.5739 & \textbf{0.0161} \\ \bottomrule
\end{tabular}
\end{table}

At the element level, we first convert the proportions of objects into a 150-dimensional vector and then measure the appearance of the two vectors.
Results of the cosine similarly show similar findings, in comparison with the results at the scene level.
Both XGradCAM (0.0161) and EigenCAM (0.0143) have the highest cosine similarities, indicating the objects that appeared in human attention heatmaps are similar to those objects in these two XAI-based heatmaps.
It implies that these two XAI models have similar patterns with human eye-level perceptions and might be more reliable in explainable objects that trigger safety perceptions.

\begin{figure}[h]
    \centering
    \includegraphics[width=\textwidth]{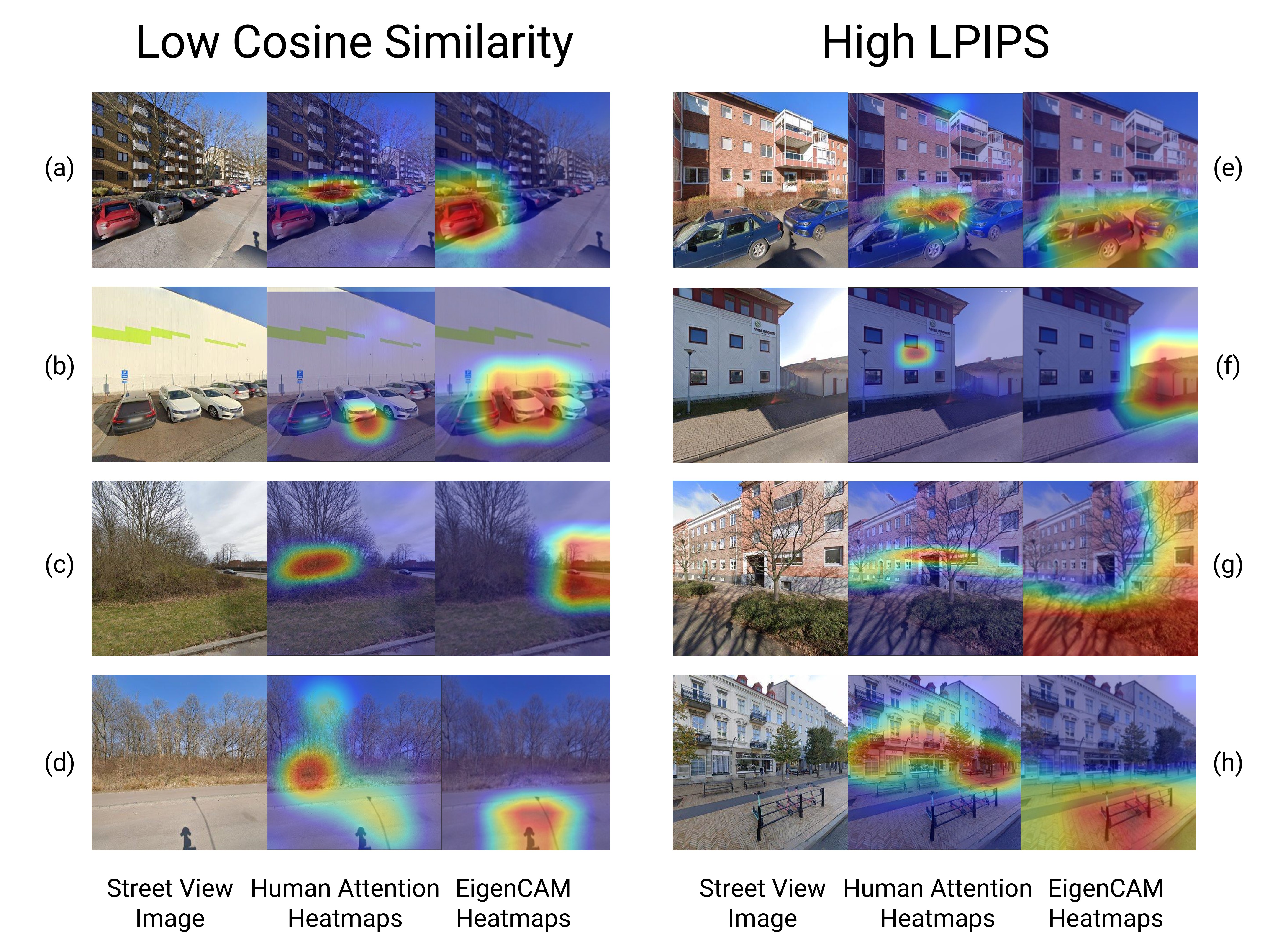}
    \caption{Representative examples of low alignment between XAI-based and human attention heatmaps. Images (a–d) show images with low cosine similarity values, and images (e–h) show images with high LPIPS values.
    }
    \label{fig:low_similarity}
\end{figure}

To gain further insight into contexts where XAI-based heatmaps diverge from human attention, we examined a few representative street view images with relatively low cosine similarity and high LPIPS values.
These images have relatively weak alignment between XAI-generated and eye-tracking heatmaps, as illustrated in Figure \ref{fig:low_similarity}. 
We selected these images from the bottom quartile of similarity scores, focusing on EigenCAM, as it shows the highest alignment across our experiments.

We observed two general patterns. 
First, while EigenCAM successfully identified objects that are associated with safety perceptions, it tended to produce broader and more diffuse heatmaps, highlighting larger regions.
In comparison, human attention maps show more selective and focused patterns, highlighting a smaller number of pixels (Figure \ref{fig:low_similarity} (a), (b), (e), (f)).
For instance, in Figure \ref{fig:low_similarity} (a), (b), and (e), EigenCAM highlighted a wide area including multiple vehicles, whereas participants tended to focus only on one or two vehicles directly related to perceived safety.
This discrepancy likely reflects the selective nature of human vision to concentrate on a small number of visual elements, whereas CAM techniques highlight more generalized regions.
These examples illustrate that low similarity may arise even when both heatmaps highlight the same objects.
Second, it is possible that the semantic focus of EigenCAM-based heatmaps is notably different from that of human observers. 
For example, in Figure \ref{fig:low_similarity} (g) and (h), participants focused on facade textures, while CAM-based heatmaps highlighted pathways.
Our findings highlight notable limitations in current XAI techniques. 
When leveraging AI for urban planning practices, models that do not align with human perception could lead to misleading conclusions. 
Therefore, we advocate for the need to develop human-aligned XAI methods that better reflect human perceptions.
Additionally, it is necessary to be cautious when interpreting XAI outputs in supporting urban decision makings.

\section{Discussions}
We discuss several takeaways from this paper below.
\subsection{Associations between Safety Perceptions and Built Environment}

This study deepens our understanding of how the built environment influences perceived safety by incorporating human gaze behavior through eye-tracking systems. In contrast to traditional computer vision approaches that treat all objects equally, our results reveal that human attention is selectively drawn to micro-level features. This aligns with the psychological principle of selective attention \citep{lavie2004load}.
We observed that urban features such as stairways, floors, and public space amenities, such as booths, flags, boxes, might be associated with safety perceptions, as they may enhance visibility and active use of places,
While vehicles and objects indicating potential barriers such as fences and hovels might be associated with negative safety perceptions.
Our findings complement and extend prior studies. For example, \citet{dong2023assessing} highlighted natural views and amenity-related features in influencing positive urban behaviors, which aligns with our findings to enhance safety perceptions.
While we have also noticed some discrepancies as sky might attract less attention, which is different from prior work \citep{ogawa2024evaluating}.
Our study contributes to the literature to move from ``what is present" to ``what people look at", offering actionable design guidance for building safer neighborhoods.

A key finding is the mismatch between the visual size of objects in street view images and the amount of attention they receive, as displayed in human attention heatmaps.
Though occupying more pixels in images, elements like the sky or ceilings attract significantly less attention.
Conversely, certain features receive disproportionate visual focus, especially those related to navigation, rest, or human activity.
These findings support classic theories in environmental design and criminology. For instance, the prominence of micro-scale elements in ``safe'' images echoes Newman's Defensible Space (\citeyear{newman1973defensible}), which emphasizes territorial cues such as thresholds and semi-private spaces that enhance perceived control and ownership. 
Similarly, our results align with principles from Crime Prevention Through Environmental Design (CPTED) \citep{jeffery1971crime}, which advocate for natural surveillance, territorial reinforcement, and spatial legibility as key strategies to reduce crime and enhance safety perception. 
The frequent association of vans and large vehicles with perceived ``unsafe'' environments reflects the logic of the Broken Windows Theory \citep{wilson1982broken}, where visual signals of disorder and transience may erode one's sense of security. 
Additionally, stairways and railings contribute to perceived safety by offering vantage points that are associated with the core Prospect-Refuge Theory \citep{appleton1975experience}. 
Symbolic or identity-linked elements, like signage and flags, illustrate how urban legibility and recognition influence perception.
These insights offer useful references for urban design and planning practices. Rather than focusing solely on large-scale openness or aesthetic uniformity, designers may consider incorporating perceptually salient features that enhance spatial legibility and support informal territorial cues. 
Especially in the case of pre-designed or existing environments, these findings could inform visual simulation-based evaluations before real-world implementation, allowing for perception-informed feedback loops.

Some limitations warrant further discussion. 
While our analysis reveals how certain physical features influence gaze behavior and perceived safety, the concept of ``safety'' may vary across contexts and individuals and is also intertwined with social and situational environments.
Despite that fear and safety perceptions are used interchangeably in this manuscript, they have different meanings.
Fear refers to an affective or emotional state, and safety perceptions reflect individuals cognitive judgments of environments. 
In our study, the question ``Which place looks safer?" may elicit more cognitive evaluations, rather than internal psychological distress. 
The interpretations of safety may vary across individuals, as someone may focus on crime-related concerns, while others may consider traffic safety or environmental hazards.
Furthermore, social contexts play an important role in shaping safety perceptions.
Factors such as the presence of other people, the topological relationships among urban elements, the perceived possibility of being seen by others, as emphasized by ``eyes on the street'' proposed by \citet{alma990001794640106761}, are crucial in shaping the sense of safety. 
Our current methodology does not fully capture these individual differences and social dimensions, and focuses primarily on visual stimuli.
Despite these limitations, our study contributes to the existing body of research on safety perceptions and place perception by introducing more human experiences with psychological methods.
To provide a more comprehensive understanding of safety perceptions in urban contexts, more behavioral data that integrates affective, cognitive, and social aspects of place perceptions is necessary in future studies.

\subsection{Advanced Technology for Environmental Perception}
This paper serves as an example of integrating emerging datasets and technology with environmental psychology to enrich our understanding of the built environment.
Using an interdisciplinary approach that combines emerging datasets and technologies, we could measure subjective human experiences at place in response to physical built environments.
On the one hand, we have demonstrated the opportunities of utilizing geospatial data science to obtain deeper insights into environmental perception studies and the complex interactions between the built environment and human perceptions.
Using street view images and deep learning approaches, we could characterize and model the built environment effectively and efficiently. 
On the other hand, the use of several advanced psychological methods, such as eye-tracking systems, opens new avenues for decoding the nature of how individuals interact with their surrounding environments.
We advocates for more human-centered insights.
By combining the two approaches together, this paper contributes to a more holistic view of environmental perception, offering a comprehensive framework that potentially could guide both research and practical applications in urban planning and public policy.

Beyond safety perceptions, the integration of advanced technologies also brings more opportunities for the future environmental perception studies.
Currently, advanced AI algorithms, like those supported by Large Language Models (LLMs), can now communicate with people in a natural way and even enrich our understanding of human subjectivity and feelings.
For instance, \citet{jang2023understanding} have leveraged LLMs to explore place identity across global cities.
Additionally, emerging psychological techniques such as fMRI (functional magnetic resonance imaging) and brain–computer interfaces have provided fresh insights into human cognition and perceptions \citep{yang2025neurocognitive, zhao2024neural}.
This study demonstrates how combining the two types of advanced technologies could facilitate our understanding of environmental perceptions.
In addition, the emergence of virtual reality (VR) and augmented reality (AR) technologies, such as the Apple Vision Pro and Meta Quest may offer new opportunities for modeling human-environment interactions \citep{yang2025neurocognitive}.
Leveraging these emerging technologies may further advance our knowledge of the digital world, beyond our physical environment.
We advocate for further interdisciplinary studies to be performed in the future to analyze the complex human-environment interactions.

Equally important, it is also necessary to acknowledge that using eye-tracking systems has challenges as well.
Due to the operating complexity, time, and cost when deploying eye-tracking systems, the sample size and geographic coverage are limited.
Therefore, a mixed-method design could significantly inform our understanding of the human-environment interactions: using geo-big data and AI models could estimate large-scale human place perceptions, and targeted eye-tracking subsamples could help uncover deeper mechanisms of the human-environment relationships.
Given that different methods have their pros and cons, it is necessary to consider their nuances when implementing them in real-world practices.

\subsection{Reliability and Ethical Implications of AI Approaches}
More importantly, we evaluated the trustworthiness of using street view images and computer vision methods for measuring human safety perceptions.
Deep learning and AI approaches have long been criticized due to their ``black-box'' nature.
Researchers have developed a series of XAI models to reveal the underlying mechanisms of AI, but the reliability of these approaches has not yet been thoroughly validated.
Alarmingly, prior studies have already treated XAI outcomes as ground truth outcomes.
However, with numerous XAI approaches based on a variety of hypotheses, which XAI model is the most trustworthy and reliable?
This paper involves human-in-the-loop comparisons between findings from eye-tracking systems and XAI models.
Through a critical analysis, we delve into the ethical considerations associated with deploying emerging technologies in urban studies. 
We not only measure the similarities between human perspectives and AI perspectives but also emphasize the significance of maintaining a human-centric approach when leveraging emerging technologies.
It is worth noting that, due to the complex process of AI models, different XAI models may have varied performances across applications and contexts.
Our study could represent a start towards encouraging more human-centered approaches to be leveraged for critically evaluating and validating the trustworthiness of the method before integrating them into urban planning practices.
This helps build trust and responsible AI applications, leading to the generation of valuable insights to enhance our understanding of safe environments and human-environment relationships more broadly.

Furthermore, the advent of deep learning could be traced back to the invention of perceptron, designed to mimic human cognitive behaviors \citep{raiko2012deep}.
Our findings indicate that there is still gaps in the capabilities of existing XAI methods to accurately infer human behaviors.
While these XAI methods can offer valuable insights, the applications of these findings need to be approached with caution before full integration into practices. 
We advocate for integrating more human-centered insights, especially from the fields of neuroscience and psychology, to facilitate the development and assessment of AI methods \citep{zhao2024neural, ye2025human}. 
Incorporating these perspectives may enhance the design of AI systems to more closely reflect actual human behaviors, thereby increasing their reliability and trustworthiness.

\subsection{Limitations and Future Directions}
There is still some room for further enhancement of the proposed experiment.
One issue refers to the sample size and geographic representativeness of the dataset.
It is resource-intensive to recruit participants to wear eye-tracking devices to collect experimental data, which limits the sample data size.
As a result, the current analysis does not support comparisons across demographic subgroups, such as gender or socioeconomic background \citep{zhou2025using}.
Also, we have only deployed studies in one city and have merely focused on safety perceptions. 
Future experiments might be performed across different areas to provide more generalized results and outcomes, and adapted to other perceptual dimensions commonly studied in urban studies, such as liveliness and beautiful.
Also, future studies may collect pre-annotated image datasets that indicates safety perceptions from other social media platforms to supplement street view images to enhance the generalizability of this study.

We also recognize the need to integrate more psychological knowledge into this study and more geography and urban studies.
There have been limited studies at the intersections between place perception studies and psychological methods.
We acknowledge that eye-tracking experiments may introduce some potential biases in measuring the ``sense of safety'' in this study.
For example, people may tend to focus on large foreground elements or independent elements \citep{zhang2017toward}.
Due to the sample size, it is challenging to analyze detailed gaze trajectories to understand topological relationships among street elements.
Integrating temporal perspectives of eye-tracking systems may offer additional insights to describe the nuanced cognitive mechanisms.
Our study did not delve into these psychological dimensions.
Thus, on the one hand, we call for more interdisciplinary collaborations to strengthen the methodological robustness of using psychological methods in future studies and to facilitate a more comprehensive integration of diverse disciplinary perspectives. 
Additionally, incorporating more human-centered approaches into geography and urban studies could enrich our nuanced understanding of the complex human-environment interactions to better modeling human subjective experiences at places. 

Another important limitation of this study is the lack of consideration for participants’ sociodemographic backgrounds. The perception of “safety” is highly subjective and may vary significantly across different population groups. For example, children and older adults may prioritize ease of mobility or visibility of assistance, while people with disabilities might focus on barrier-free access and navigability. 
Individuals of different genders and from diverse ethnic or cultural backgrounds may interpret environmental cues differently due to prior experiences or systemic inequalities \citep{zhou2025using}.
Moreover, the degree of familiarity with the environment, such as whether a person is a local resident or a visitor, can also influence safety perception. Residents may feel more at ease in spaces they know well, even if those spaces appear disordered to outsiders. 
In our study, over 93\% of participants identified as White, such demographic homogeneity may limit our ability to generalize findings to more diverse populations. 
Given that our study did not systematically account for these variables, future research may aim to incorporate a more diverse and representative sample to better understand how different demographic (e.g., gender) and social groups experience safety within the built environment.

\section{Conclusions}
In conclusion, by integrating eye-tracking systems and street view images, we examined human safety perceptions within the urban environment comprehensively.
Several objects in urban environments were identified that draw visual attention and affect human safety perceptions when observing the built environment.
We discovered that certain urban infrastructure elements, such as stairways and floors, and public space features, such as flags and chairs attract more human attention when determining safe environments, while the presence of vehicles might be associated with unsafe environments.
Open-space features, such as sky, receive less attention despite their high proportions in images.
More importantly, by comparing heatmaps generated from eye-tracking systems and those generated using eXplainable AI (XAI) techniques, we evaluate the trustworthiness of these machine-based metrics in measuring safety perceptions.
We suggest that the results generated by XGradCAM and EigenCAM are the most aligned with human safety perceptions in a Swedish context.
This study demonstrates how integrating emerging technologies including street view images, deep learning, and eye-tracking systems could deepen our knowledge of human-environment relationships, and could inform the future design of safe environments and communities.


\bibliographystyle{apacite}
\bibliography{reference}

\section{Appendix}
\subsection{Dataset Construction}
\label{Appendix:dataset}

We constructed our sample dataset by selecting 300 street view images among all street view images in Helsingborg, Sweden, to represent varying urban landscapes.
To ensure broader geographic representation across the city, we performed a spatially random sampling of street view images. Following this, we manually reviewed and refined the sample by removing images that were either completely black or significantly distorted. We also removed a few images depicting duplicated scenes to enhance the diversity and representativeness of visual contexts in the experiment. 
Both the global model and the Stockholm model were applied to all images, yielding continuous safety perception ratings on a 1–9 scale.

We then categorized the images into three safety perception levels based on these scores \citep{tang2024urban, lei2024evaluating}: \textit{high}, \textit{medium}, and \textit{low}.
Adopting such a ternary classification rather than directly using continuous safety perception scores (1-9) was intended to mitigate the following two potential issues: (1) subtle differences in a continuous scoring scale may not be perceptually significant to participants; and (2) to reduce the influences on individual differences in eye-tracking experiments \citep{valuch2015using}.
Each category contains 100 street-view images to be utilized in our further experiments.
Specifically, images classified within the top 20\% according to both scores were designated as having \textit{high safety} perceptions, whereas those in the bottom 20\% were assigned as having \textit{low safety} perceptions.
Images that fell between the 40\% and 60\% percentiles were considered to have \textit{medium safety} perceptions. 
We excluded images in the 20–40\% and 60–80\% ranges to avoid ambiguous stimuli and to maintain clear perceptual contrasts among categories.
Upon creating these categories, we randomly selected 100 images from each safety perception category and compiled a balanced dataset of 300 images.
It should be noted that the three-category classification of safety perception scores (high, medium, low) was applied solely as a pre-processing step to construct a balanced sample dataset for subsequent analysis. 
Such a categorization does not imply a direct or absolute relationship between the assigned score and the safety perceptions evoked in later user studies. 
An image with a high safety perception score does not necessarily elicit strong feelings of safety from participants.
This demonstrates the importance of keeping a human-in-the-loop approach, where human behavior responses are important rather than relying solely on algorithmic predictions.
The number of selected images was determined by our preliminary estimates of the number of participants in our study, aiming to strike a balance between the frequency of image occurrences and the number of participants.
This consideration ensures that most images were compared with other images over five times, thereby facilitating more reliable estimations of human safety perceptions.
This dataset will be further employed in the subsequent survey to collect human safety perceptions with eye-tracking systems in Section \ref{section:survey}.


\subsection{Image Similarity Metrics}
\label{Appendix:similarity}
We offer more technical details about the L2 loss and LPIPS scores leveraged for measuring image similarities in Section \ref{image_similarity}.

The L2 loss is computed as the Euclidean distance between the corresponding pixel values $p(i,j)$ of two images $m$ and $n$. 
\begin{equation}
L2\text{ loss} = \sqrt{\sum_{i=1}^{h}\sum_{j=1}^{w} (p_{m}(i,j) - p_{n}(i,j))^2}
\end{equation}

While the L2 loss has been commonly used as a basic metric for evaluating image similarities, it may not fully represent the perceived similarity because it is sensitive to individual pixel matches. 
Furthermore, it is challenging to understand the meanings of such error representation metrics.
Thus, we utilized the LPIPS score, a validated metric for measuring perceptual similarity between image pairs in prior studies, to evaluate the similarity between human attention heatmaps and XAI-based heatmaps \citep{zhang2018unreasonable, jang2024place}.
LPIPS quantifies the Euclidean distance between feature vectors of images extracted from a pretrained deep convolutional neural network, like AlexNet \citep{krizhevsky2012imagenet}. 
We also applied this network as a feature extractor to calculate LPIPS scores for comparing two images. 
By combining these two measures, we measured the image similarity between the human attention and XAI-based heatmap images at the scene level.
Results help identify the XAI model that best matches human visual attention.

\end{document}